\documentclass[ngerman,english]{article}
\usepackage[T1]{fontenc}
\usepackage[latin9]{inputenc}
\usepackage{float}
\usepackage{mathtools}
\usepackage{amssymb}
\usepackage{graphicx}

\makeatletter

\providecommand{\tabularnewline}{\\}
\newcommand{\lyxdot}{.}

\newcommand{\lyxaddress}[1]{
\par {\raggedright #1
\vspace{1.4em}
\noindent\par}
}

\makeatother

\usepackage{babel}
\begin{document}

\title{Improved Time of Arrival measurement model for non-convex optimization
with noisy data}

\author{Juri Sidorenko, Leo Doktorski, \\
Volker Schatz, Norbert Scherer-Negenborn, Michael Arens}
\maketitle

\lyxaddress{Fraunhofer IOSB, Ettlingen Germany \\
juri.sidorenko@iosb.fraunhofer.de, Tel.: +49 7243 992-351}


\section{Abstract}

The quadratic system provided by the Time of Arrival technique can
be solved analytical or by optimization algorithms. In real environments
the measurements are always corrupted by noise. This measurement noise
effects the analytical solution more than non-linear optimization
algorithms. On the other hand it is also true that local optimization
tends to find the local minimum, instead of the global minimum. This
article presents an approach how this risk can be significantly reduced
in noisy environments. The main idea of our approach is to transform
the local minimum to a saddle point, by increasing the number of dimensions. 

\section{Introduction}

In position estimation the Time of Arrival (ToA) \cite{TOA} technique
is standard. The area of application extends from satellite based
systems like GPS \cite{GPS1}, GLONASS \cite{Glonass}, Galileo \cite{Galileo},
mobile phone localization (GSM) \cite{GSM}, radar based systems such
as UWB \cite{UWB}, FMCW radar \cite{Radar} to acoustic systems \cite{Accustic}.
\\
The ToA technique leads to a quadratic equation. Optimization algorithms
used to solve this system depends on the initial estimate. Unfortunately
chosen initial estimates can increase the probability to convergence
to a local minimum. In some cases it is possible to transform the
quadratic to a linear system \cite{ION_JS,Sidorenko2016,Hmam2010}.
This linear system can be used to provide an initial estimate. On
the other hand, the linear system is more affected by noise, compared
to the quadratic system \cite{ION_JS,Sidorenko2016}. In practice,
a combination of both methods is used to obtain the unknown position
of an object. \cite{Abel1991,Bancroft1985,Chaffee1994}. However,
the initial estimates by a linear solution only applies if the base
station positions are known. This article presents an approach how
the risk of convergence to a local minimum during the optimization
process can be significantly reduced for the ToA technique. The approach
does not require initial estimations provided by a linear solution,
rather the insertion of an additional variable is used to transform
a local minimum to a saddle point at the same coordinates. In order
to simplify the prove of our approach, it is assumed that the position
of the base stations are known. Our approach was inspired by dimension
lifting \cite{Balas2005,Lifting,Lift} and concave programming \cite{Liu1995}.
Dimension lifting introduces an additional dimension to transform
a non-convex to a convex feasible region. Concave programming describes
a non-convex problem in terms of d.c. functions (differences of convex
functions). In our method, the non-convex problem remains non-convex.
In the publication \cite{Sidorenko2018} it was shown that this approach,
reduces the risk of convergence to a local minimum for measurements
without noise. This elaboration is more focused on the effect of noise
on our approach.\\

This paper is organized as follows. The third section, introduces
the objective functions $F$ and the corresponding improved objective
functions $F_{L}$. In Section four, we use Levenberg-Marquardt algorithm
\cite{Levenberg-Marquardt} to illustrate the optimization steps for
$F$ and $F_{L}$. The last section address the results of the optimization
algorithm with randomly selected constellations and different amounts
of noise. 

\section{Methodology}

\begin{table}[H]
\begin{centering}
\caption{Used notations\label{tab:Exmplation-of-the}}
\ \\
\par\end{centering}
\centering{}%
\begin{tabular}{|c|c|}
\hline 
Notations & \selectlanguage{ngerman}%
Definition\selectlanguage{english}%
\tabularnewline
\hline 
\hline 
$x,y,z$ & Estimated position of the transponder $T$ \tabularnewline
\hline 
$x_{G},y_{G},z_{G}$ & Ground truth position of the transponder T\tabularnewline
\hline 
$a_{i},b_{i},c_{i}$ & Ground truth position of base stations $B_{i}$, $1\leq i\leq N$\tabularnewline
\hline 
$d_{i}$ & Distance measurements between $B_{i}$ and $T$\tabularnewline
\hline 
$\lambda$ & Additional variable\tabularnewline
\hline 
\end{tabular}
\end{table}

\begin{figure}[H]
\centering{}\includegraphics[scale=0.5]{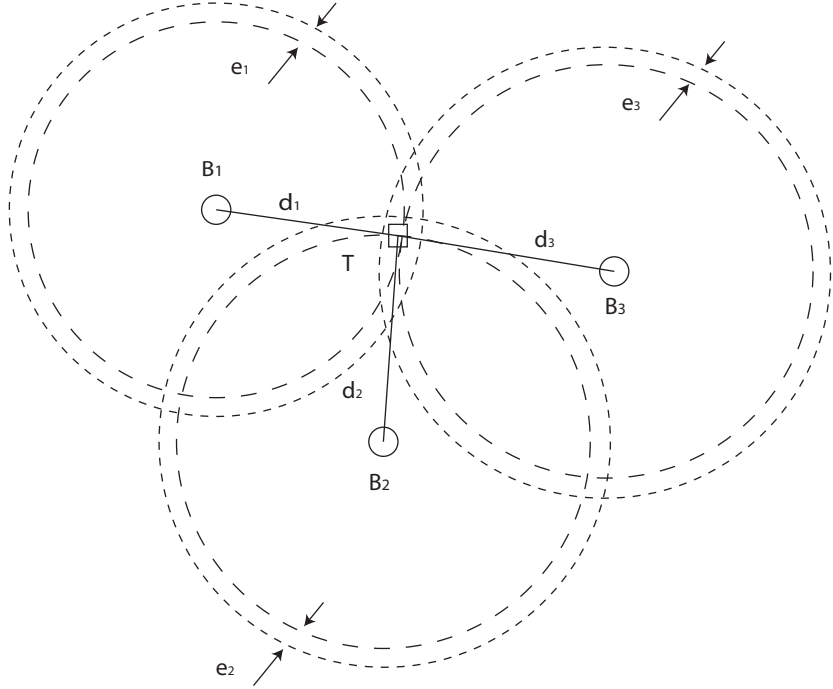}\caption{The dashed circles with a smaller radius are the true distances between
base stations $B_{i}$ and transponder $T$. The dashed circles with
a radius of $d_{i}+e_{i}$ are the false measurements due to noise.
\label{fig:intro} }
\end{figure}

Figure \ref{fig:intro} shows three base stations $B_{i}$ at known
positions $(a_{i},b_{i},c_{i})$, and one transponder $T$ at unknown
position $(x,y,z)$. The distances measurements $d_{i}$ between base
stations $B_{i}$ and the transponder $T$ are known. The unkown position
of the transponder $T$ can be estimated by the known positions of
the base stations $B_{i}$ and the distance measurements $d_{i}$.
This data is effected by gaussian noise $e_{i}$. 

\subsection{Mathematical formulation}

The distance measurements between the base stations $B_{i}$ and transponder
$T$ are defined as

\begin{center}
\begin{tabular}{ccc}
$d_{i}=\sqrt{(x_{G}-a_{i})^{2}+(y_{G}-b_{i})^{2}+(z_{G}-c_{i})^{2}}.$ &  & $1\leq i\leq N$\tabularnewline
\end{tabular}
\par\end{center}

Unknown position of transponder T can be found by solving eq. (\ref{eq:obj1}). 
\begin{itemize}
\item Objective function one:
\end{itemize}
\begin{equation}
F_{1}(x,y,z)\coloneqq\sum_{i=1}^{N}\left[\sqrt{(x-a_{i})^{2}+(y-b_{i})^{2}+(z-c_{i})^{2}}-d_{i}+e_{i}\right]^{2}\label{eq:obj1}
\end{equation}
The solving of eq.(\ref{eq:obj1}) can be done by non-convex optimization
\cite{LeastSquares} $F_{i}(x,y,z)\rightarrow argmin$. Alternatively,
the non-linear system can be transformed into a linear system \cite{ION_JS,Sidorenko2016}.
In more complex cases where the positions of base stations $B_{i}$
are unknown this is not possible at all. With regard to future extensions
to determining the base station positions as well as the location
of the transponder $T$, this article focuses on finding a solution
with a non-convex optimization algorithm.

\subsection{Reason for the approach}

The objective function (\ref{eq:obj1}) is non-linear and non-convex.
The optimization of the objective function can cause to convergence
to a local minimum $L$ instead the global minimum $G$ (see Table
\ref{tab:Exmplation-of-the}). In our approach instead the $F_{1}$
the improved objective function $F_{L1}$ is used. This function has
an additional variable $\lambda$ compared to the function $F_{1}$. 
\begin{itemize}
\item Improved objective function one:
\end{itemize}
\begin{equation}
F_{L1}(x,y,z,\lambda)\coloneqq\sum_{i=1}^{N}\left[\sqrt{(x-a_{i})^{2}+(y-b_{i})^{2}+(z-c_{i})^{2}+\lambda^{2}}-d_{i}+e_{i}\right]^{2}\label{eq:IOF1}
\end{equation}
In \cite{Sidorenko2018} we have proven that the improved objective
function two $F_{L2}(x,y,z,\lambda)\coloneqq\sum_{i=1}^{N}\left[(x-a_{i})^{2}+(y-b_{i})^{2}+(z-c_{i})^{2}+\lambda^{2}-d_{i}^{2}\right]^{2}$
with an additional variable, transforms the local minimum to a saddle
point at $\lambda=0$. Furthermore it was shown that no further local
minima exist for $\lambda\neq0$ at non trivial constellations. The
same effect was demonstrated numerically for eq.(\ref{eq:obj1}) and
eq.(\ref{eq:IOF1}). The final proof of the hypothesis was provided
with the help of the Cauchy-Bunyakovsky-Schwarz inequality\cite{SchwarzUngleichung}.
Alternatively, the equation 29 in \cite{Sidorenko2018} can also be
obtained from the variance $Var(X)=\mathbb{E}\left(\left(X-\mathbb{E}\left(X\right)\right)^{2}\right)=\mathbb{E}\left(X\right)^{2}-\left(\mathbb{E}\left(X\right)\right)^{2}$.
The base stations should have a variance in position higher or equal
to zero $0\leq Var(\left\{ a_{i}\right\} )=\frac{1}{N}\sum_{i=1}^{N}a_{i}^{2}-\left(\frac{1}{N}\sum_{i=1}^{N}a_{i}\right)^{2}=\frac{1}{N}\sum_{i=1}^{N}a_{i}^{2}-\frac{1}{N^{2}}\left(\sum_{i=1}^{N}a_{i}\right)^{2}$.
This leads to the final term $\sum_{i=1}^{N}a_{i}\leq\sqrt{N}\sqrt{\sum_{i=1}^{N}a_{i}{}^{2}}$.
In this article the measurement data is effected by noise, therefore
the objective function one (\ref{eq:obj1}) is used. In contrast to
objective function two is this function statistically correct in presence
of noise.

\subsection{Two dimensional example }

\begin{figure}[H]
\begin{centering}
\includegraphics[scale=0.4]{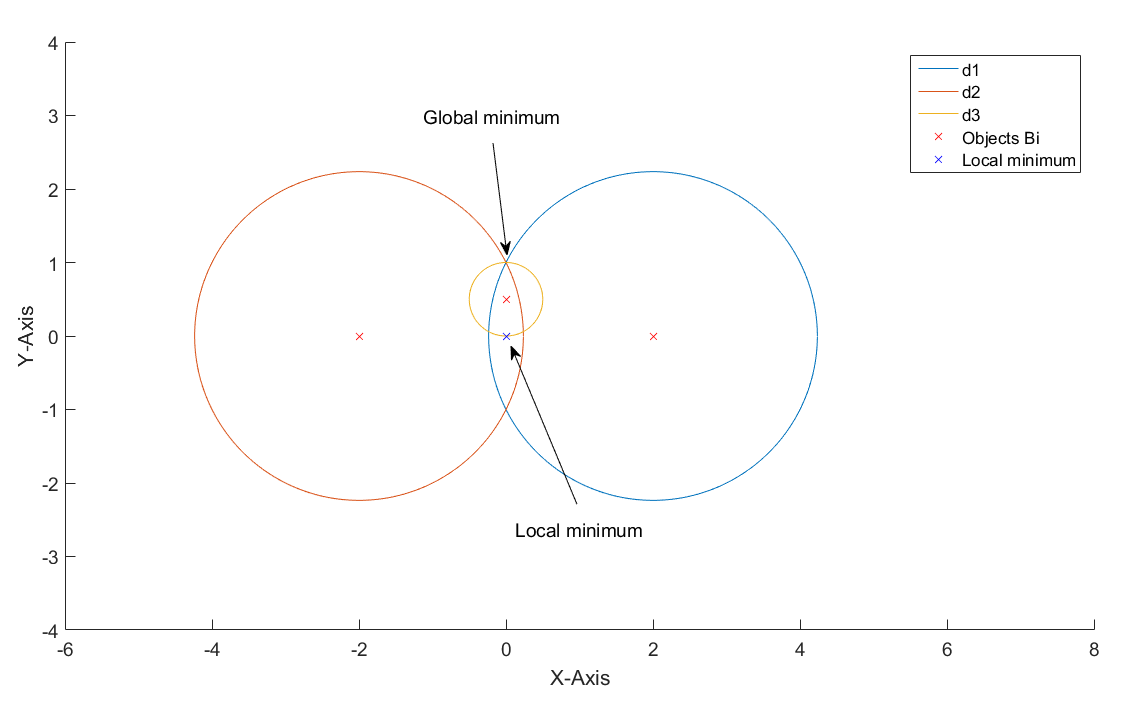}
\par\end{centering}
\caption{The circles represents the true distance between base stations $B_{i}$
and the global minimum. Blue circle is the distance between base station
$B_{2}$ and transponder. Red circle the distance between base station
$B_{3}$ and transponder. Yellow circle is the distance between base
station $B_{1}$ and transponder .\label{fig:F:-Objective-function-1}}
\end{figure}

\begin{figure}[H]
\begin{centering}
\includegraphics[scale=0.4]{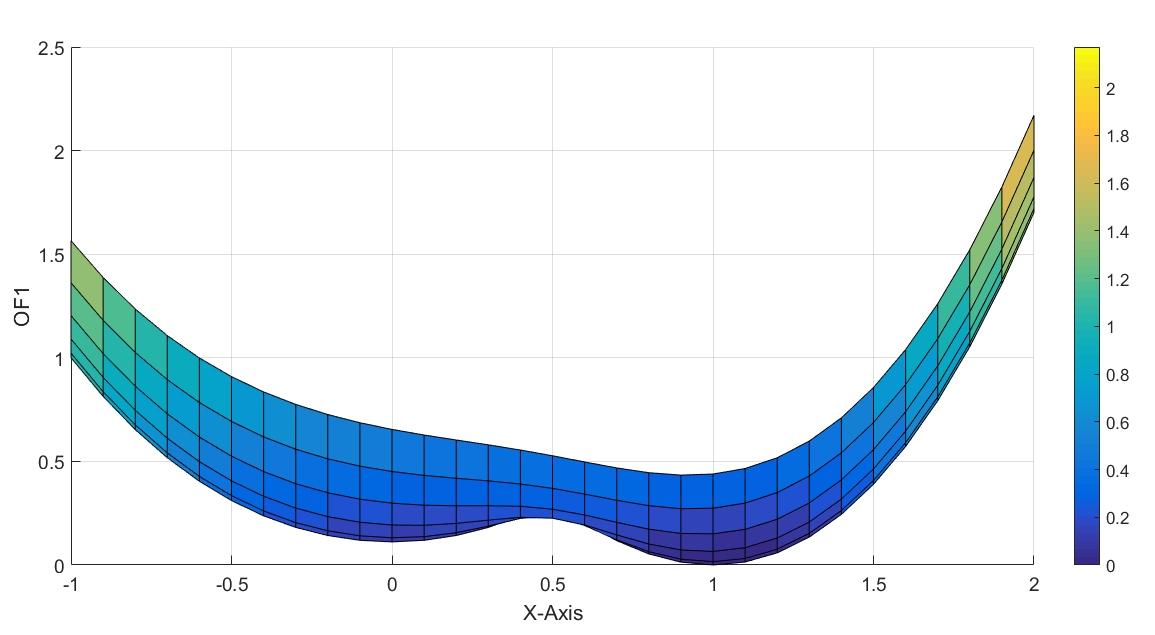}
\par\end{centering}
\caption{Local minimum at L(0,0) and global optima at G(1,0). Colors from blue
to yellow showing the result of the objective function \label{fig:Local-minima-at}}
\end{figure}

In this section an example is created with known coordinates of the
global at $G(1,0)$, local minimum $L(0,0)$ and no noise. This example
has the aim to illustrate the converging steps of the Levenberg-Marquardt
algorithm for the $F_{1}$ and $F_{L1}$. The positions of the local
and global minimum leads to the coordinates of base stations $B_{i}$
(See Table \ref{tab:Coordinates-of-objects} ). Figure \ref{fig:F:-Objective-function-1},
shows the coordinates of base stations $B_{i}$, which are located
in the center of the circles. And the figure \ref{fig:Local-minima-at},
presents the search space of objective function $F_{1}$. 

\begin{table}[H]
\caption{Coordinates of base stations $B_{i}$\label{tab:Coordinates-of-objects}}
\ \\
\centering{}%
\begin{tabular}{|c|c|c|}
\hline 
Base stations  & X-Position & Y-Position \tabularnewline
\hline 
\hline 
$B_{1}$ & $0.5$ & $0$\tabularnewline
\hline 
$B_{2}$ & $0$ & $2$\tabularnewline
\hline 
$B_{3}$ & $0$ & $-2$\tabularnewline
\hline 
\end{tabular}
\end{table}

\subsubsection{Local optimization}

The Levenberg-Marquardt algorithm uses the derivative to obtain the
stepsize, therefore it is important that the initial estimate for
the additional variable $\lambda$ is non-zero. Otherwise $\lambda$
remains zero, and $F_{L1}$ is effectively reduced to $F_{1}$.

Table \ref{tab:Initial-position} shows initial estimates of the optimization.

\begin{table}[H]
\caption{Iteration steps of the Levenberg-Marquardt for $F_{1}$and $F_{L1}$.\label{tab:Initial-position}}
\ \\
\centering{}%
\begin{tabular}{|c|c|c|c|}
\hline 
 & $x$ & $y$ & $\lambda$\tabularnewline
\hline 
\hline 
Initial estimate & 2 & -1 & 1\tabularnewline
\hline 
\end{tabular}
\end{table}

\begin{figure}[H]
\begin{centering}
\includegraphics[scale=0.5]{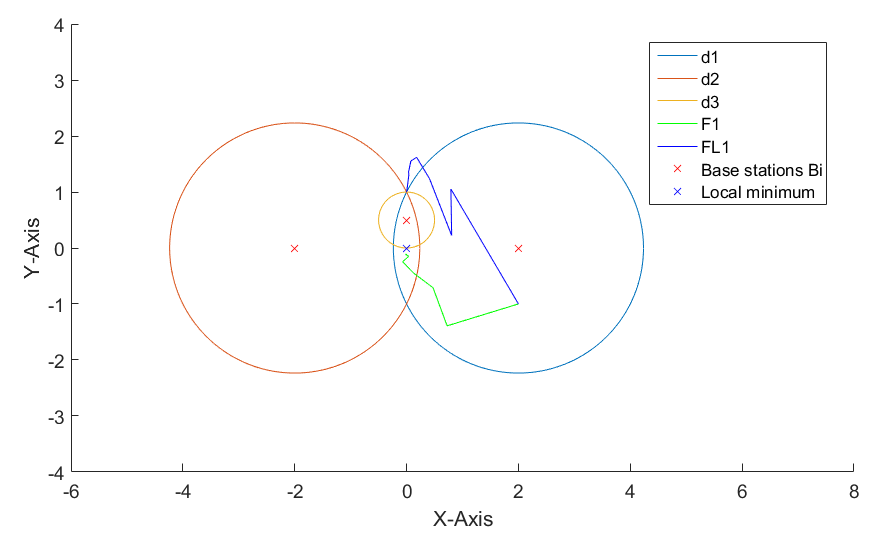}
\par\end{centering}
\caption{Iteration steps of the Levenberg-Marquardt for $F_{1}$and $F_{L1}$.
F1: Objective function $F_{1}$. FL1: Improved objective function
one $F_{L1}$. Blue line: Opimization steps, Gree line: Optimization
steps improved function. The circles blue, bed and yellow are the
distances between base stations $B_{i}$ and transponder $T$ \label{fig:F:-Objective-function}}
\end{figure}

In Figure \ref{fig:Local-minima-at} the result of the optimization
can be observed. The blue path shows the steps of the improved objective
function $F_{L1}$, which converge to the global minimum $G(1,0)$.
On the other hand, the original objective function $F_{1}$ represented
by the green line, converges to the local minimum  $L(0,0)$.\\
If the measurement is effect by noise, the residues would be higher
than zero at the global minimum. With more additional variables (eq.
\ref{eq:moreVariables}) the error splits up between the additional
variables in the manner $\lambda=\sqrt{\sum_{i=1}^{N}\lambda_{i}{}^{2}}$. 

\begin{equation}
F_{L1}(x,y,z,\lambda_{1})\coloneqq\sum_{i=1}^{N}\left[\sqrt{(x-a_{i})^{2}+(y-b_{i})^{2}+(z-c_{i})^{2}+\lambda_{1}{}^{2}}-d_{i}-e_{i}\right]^{2}
\end{equation}

\begin{equation}
F_{L2}(x,y,z,\lambda_{2},\lambda_{3})\coloneqq\sum_{i=1}^{N}\left[\sqrt{(x-a_{i})^{2}+(y-b_{i})^{2}+(z-c_{i})^{2}+\lambda_{2}{}^{2}+\lambda_{3}{}^{2}}-d_{i}-e_{i}\right]^{2}\label{eq:moreVariables}
\end{equation}

We assume that the proven hypothesis \cite{Sidorenko2018} for the
improved objective function two apply as well for the improved objective
function one eq.(\ref{eq:secondDiff_smaller_zero-1}).

\begin{equation}
\left(\frac{\partial^{2}}{\partial\lambda^{2}}F_{L1}\right)(0,0,0,0)=\sum_{i=1}^{N}\frac{\left(\sqrt{a_{i}{}^{2}+b_{i}{}^{2}}-d_{i}+e_{i}\right)}{\sqrt{a_{i}{}^{2}+b_{i}{}^{2}}}<0\label{eq:secondDiff_smaller_zero-1}
\end{equation}

\section{Numerical results}

The tests were carried out with MALTAB Levenberg-Marquardt algorithm
at default settings (Table. \ref{tab:Default-Matlab-'Levenberg}).
\begin{table}[H]
\caption{Default MATLAB 'Levenberg Marquardt algorithm' parameter \label{tab:Default-Matlab-'Levenberg}}
\ \\
\centering{}{\footnotesize{}}%
\begin{tabular}{|c|c|}
\hline 
 & {\footnotesize{}Value}\tabularnewline
\hline 
\hline 
{\footnotesize{}Maximum change in variables for finite-difference
gradients} & {\footnotesize{}Inf}\tabularnewline
\hline 
{\footnotesize{}Minimum change in variables for finite-difference
gradients } & {\footnotesize{}0}\tabularnewline
\hline 
{\footnotesize{}Termination tolerance on the function value} & {\footnotesize{}1e-6}\tabularnewline
\hline 
{\footnotesize{}Maximum number of function evaluations allowed} & {\footnotesize{}100{*}numberOfVariables}\tabularnewline
\hline 
{\footnotesize{}Maximum number of iterations allowed} & {\footnotesize{}400}\tabularnewline
\hline 
{\footnotesize{}Termination tolerance on the first-order optimality} & {\footnotesize{}1e-4 }\tabularnewline
\hline 
{\footnotesize{}Termination tolerance on x} & {\footnotesize{}1e-6}\tabularnewline
\hline 
{\footnotesize{}Initial value of the Levenberg-Marquardt parameter} & {\footnotesize{}1e-2}\tabularnewline
\hline 
\end{tabular}{\footnotesize \par}
\end{table}
 The base stations $B_{i}$, transponder $T$ and initial estimates
were randomly generated in a 10x10x10 cube. Unfavorable constellation
close to collinearity have been avoided by the requirement that every
normalized singular value of the covariance matrix should be higher
than $0.1$. 
\begin{itemize}
\item Error term:
\end{itemize}
\begin{equation}
E=\sum_{j=1}^{M}\sqrt{(x-x_{G})^{2}+(y-y_{G})^{2}+(z-z_{G})^{2}}\label{eq:errorTerm}
\end{equation}

\subsection{Results of the objective function and the improved objective function\label{subsec:ResultsSection}}

\begin{figure}[H]
\begin{centering}
\includegraphics[scale=0.45]{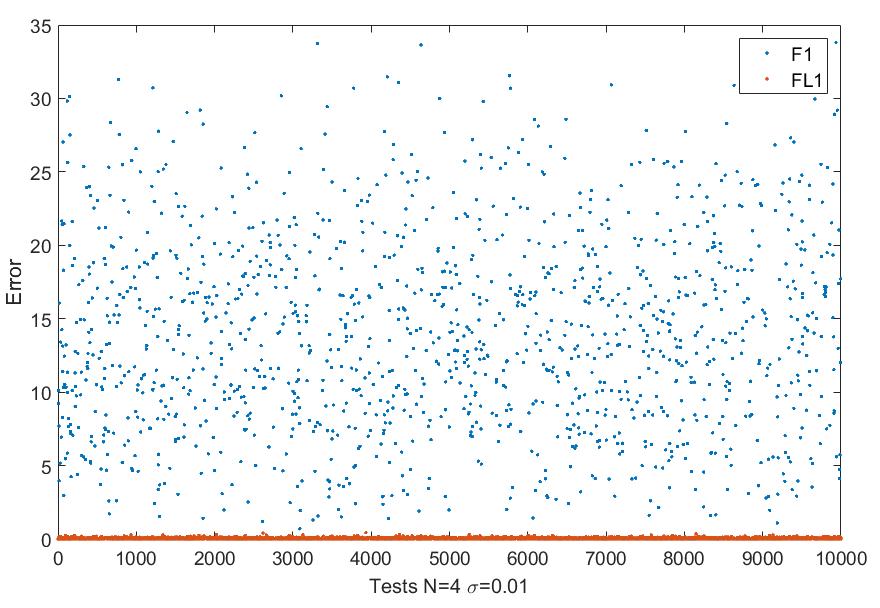}
\par\end{centering}
\caption{Blue dots: Objective function $F_{1}$. Red dots: Improved objective
function $F_{L1}$\label{fig:results1-1}}
\end{figure}

In the following section the results of the optimization with a two
dimensional $F_{1}$ are presented. Figure \ref{fig:results1-1} shows
the error term with different constellations of the four base stations
$B_{i}$. It can be seen that $F_{L1}$ has no outlier for measurement
noise smaller than $\sigma\leq0.01$. The measurement noise $e_{i}$effects
eq. (\ref{eq:secondDiff_smaller_zero-1}) and could lead to a local
minima $\left(\frac{\partial^{2}}{\partial\lambda^{2}}F_{L1}\right)(0,0,0,0)>0$.
Therefore, with higher noise are convergences to local minima also
possible for the improved objective function. 

\begin{table}[H]
\centering{}\caption{The examples are based on a 2-D model with 4 base stations $B_{i}$.
$F_{1}$: Objective function one, $F_{L1}$: Improved objective function
one, M: Mean error, Sigma: Standard deviation, L: Amount of local
minima (Error bigger then $0.5$) \label{tab:results}. }
\ \\
\begin{tabular}{|c|c|c|c|c|}
\hline 
Noise $\sigma$ & Objective function &  $M\pm\sigma$ & L & $M\pm\sigma$ without outlier\tabularnewline
\hline 
\hline 
0.01 & $F_{1}$ & $1.9139\pm5.3541$ & 1357 & $0.0344\pm0.0286$\tabularnewline
\hline 
0.01 & $F_{L1}$ & \textbf{$0.0357\pm0.0304$} & 0 & $0.0357\pm0.0304$\tabularnewline
\hline 
\hline 
0.05 & $F_{1}$ & $1.8155\pm5.0454$ & 1313 & $0.1306\pm0.0810$\tabularnewline
\hline 
0.05 & $F_{L1}$ & \textbf{$0.1746\pm0.1505$} & 362 & $0.1542\pm0.0986$\tabularnewline
\hline 
\hline 
0.1 & $F_{1}$ & $1.9939\pm5.1490$ & 1900 & $0.2250\pm0.1133$\tabularnewline
\hline 
0.1 & $F_{L1}$ & \textbf{$0.3426\pm0.2920$} & 1743 & $0.2419\pm0.1191$\tabularnewline
\hline 
\end{tabular}\\
\end{table}

The mean error without the outlier is higher for the improved objective
function one, due to the fact that with more dimensions the ratio
between the number of equations with respect to the amount of unknown
dimensions and is decreasing.

\subsection{Results with more than one additional variable}

\begin{figure}[H]
\begin{centering}
\includegraphics[scale=0.45]{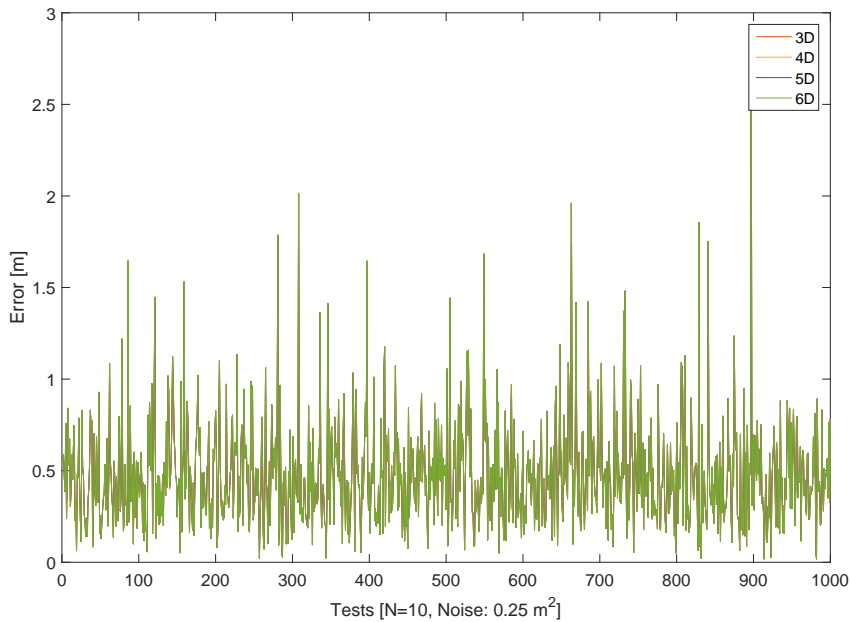}
\par\end{centering}
\caption{Tests with 10 base stations and different number of additional variables.
\label{fig:FurtherDimensions}}
\end{figure}

In figure \ref{fig:FurtherDimensions} the results with more than
one additional variable can be observed. In contrast to the results
of section \ref{subsec:ResultsSection}, all possible constellations
have been used for the lateration. Therefore, in some cases the optimization
converges to a local minimum. Regardless the number of additional
variables the results are the same, hence it makes no sense to use
more than one additional variable.

\subsection{Results with restart\label{subsec:Results-with-restart}}

\begin{figure}[H]
\begin{centering}
\includegraphics[scale=0.45]{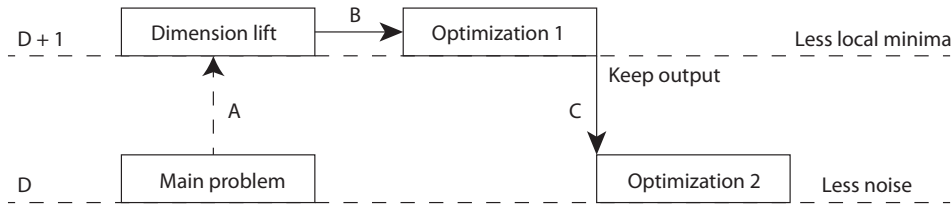}
\par\end{centering}
\caption{Flow of the optimization process. D: Optimization with the exact number
of dimensions of the model. D+1: Optimization with an additional dimension.
A,B,C order of the flow \label{fig:FlowChart}}
\end{figure}

The improved objective function $F_{L1}$ has the advantage that it
is less effected by local minima. The general objective function has
with less dimensions a better noise compensation. Therefore, it makes
sense to combine the strength of both functions. In figure \ref{fig:FlowChart}
we present a method how both effects can be used. 

At the beginning of the optimization process the objective function
is getting increasing by one additional variable $\lambda$ (step
A). In the next step B the optimization is done with the additional
variable to minimize the risk to find a local minimum. In step C,
the outcome of the optimization is used as initial estimate for the
next optimization without the additional variable. Table \ref{table_restart}
shows the results of the optimization process with restart. The number
of found outlier and mean error is smaller compared to the objective
function and the improved objective function.

\begin{table}[H]
\centering{}\label{table_restart}\caption{The examples are based on a 2-D model with 4 base stations $B_{i}$
. $F_{1}$: Objective function one, $F_{L1}$: Improved objective
function one. $F_{L1}$to $F_{1}$: Restart of the optimization $F_{L1}$
with initial estimates obtained from optimization with $F_{1}$. M:
Mean error, Sigma: Standard deviation, L: Amount of local minima (Error
bigger then $0.5$) }
\ \\
\begin{tabular}{|c|c|c|c|}
\hline 
Noise &  $M\pm\sigma$ &  L & Noise $\sigma$ without outlier\tabularnewline
\hline 
\hline 
0.01 & $0.0260\pm0.0175$ & 0 & $0.0260\pm0.0175$\tabularnewline
\hline 
0.05 & $0.1302\pm0.0900$ & 61 & $0.1269\pm0.0769$\tabularnewline
\hline 
0.1 & $0.2582\pm0.1924$ & 697 & $0.2249\pm0.1130$\tabularnewline
\hline 
\end{tabular}\\
\end{table}

\section{Discussion}

The presented method shows a huge advantage over the cassic objective
function. In \cite{Sidorenko2018} we have proven that the improved
objective function two $F_{2}$ has a saddle point at the local minimum
of objective function two $F_{2}$. In test scenraios with no or small
noise the improved objective function onw $F_{L1}$ never converges
into a local minimum. With increasing noise does the improved objective
function one $F_{L1}$ lose its ability to avoid local minima. However,
the amount of fase converages was ten times lower of $F_{L1}$ compared
to $F_{1}$. On the other side, the function $F_{1}$has a better
noise dumping than $F_{L1}$. This is due to a better ratio between
number of equations to unkown dimensions. The presented method in
section \ref{subsec:Results-with-restart} shows that this disadvange
can be overcome with a restart of the optimization with $F_{1}$with
inistial estimates provided by $F_{L1}$. In any case, it is not necessary
to implement more than one additional variable. It is important that
the initial estimate of the additional variable is unequal zero. Otherwise
gradient-based optimization algorithms like Levenberg-Marquardt would
not converge to the additional dimension. In all test scenarios the
positions of base stations $B_{i}$were known. Under the following
conditions it is also possible to obtain the solution analytically.
In the case of unknown positions of base stations $B_{i}$ and transponders
$T_{j}$ it is not feasible anymore. At this point, our approach becomes
extremely valuable.

\bibliographystyle{plain}
\bibliography{bib_dimension}

\end{document}